\documentclass{aa}
\usepackage{graphicx}

\usepackage{txfonts}

\newcommand{\cmcub}{~cm$^{-3}$}

\newcommand{\qh}{$Q({\rm{H^{0}}})$}

\newcommand{\nh}{$n({\rm{H}})$}

\newcommand{\Te}{$T_{\rm{e}}$}

\newcommand{\TrOiii}{$T_{\rm{r}}$[O~{\sc iii}]}
\newcommand{\TrOii}{$T_{\rm{r}}$[O~{\sc ii}]}
\newcommand{\TrNii}{$T_{\rm{r}}$[N~{\sc ii}]}
\newcommand{\TlOiii}{$T_{\rm{l}}$[O~{\sc iii}]}
\newcommand{\TlOii}{$T_{\rm{l}}$[O~{\sc ii}]}
\newcommand{\TlNii}{$T_{\rm{l}}$[N~{\sc ii}]}

\newcommand{\TOpp}{$T$(O$^{++}$)}

\newcommand{\TNp}{$T$(N$^{+}$)}

\newcommand{\Teff}{$T_{\rm{eff}}$}

\newcommand{\Hb}{\ifmmode {\rm H}\beta \else H$\beta$\fi}

\newcommand{\hii}{H~{\sc ii}}

\newcommand{\Hei}{He~{\sc i} $\lambda$5876}

\newcommand{\Niit}{[N~{\sc ii}] $\lambda$5755}

\newcommand{\Oii}{[O~{\sc ii}] $\lambda$3727}
\newcommand{\Oiit}{[O~{\sc ii}] $\lambda$7325}

\newcommand{\Oiii}{[O~{\sc iii}] $\lambda$5007}
\newcommand{\Oiiit}{[O~{\sc iii}] $\lambda$4363}
\newcommand{\oiii}{[O~{\sc iii}]}
\newcommand{\neii}{[Ne~{\sc ii}]}
\newcommand{\Neiii}{[Ne~{\sc iii}] $\lambda$3869}
\newcommand{\neiii}{[Ne~{\sc iii}]}

\newcommand{\Siiit}{[S~{\sc iii}] $\lambda$6312}

\newcommand{\Oiitoneb}{[O\,{\sc{ii}}]\,$\lambda$7320,7330}

\newcommand{\Oiiitoneb}{[O\,{\sc{iii}}]\,$\lambda$88\,$\mu$m}

\newcommand{\rOiit}{[O~{\sc ii}] $\lambda$3726,3729/7320,7330}
\newcommand{\rOiii}{[O~{\sc iii}] $\lambda$4363/5007}
\newcommand{\rNii}{[N~{\sc ii}] $\lambda$5755/6584}

\newcommand{\rSii}{[S~{\sc ii}] $\lambda$6731/6717}

\newcommand{\Hp}{H$^{+}$}

\newcommand{\Hep}{He$^{+}$}
\newcommand{\Hepp}{He$^{++}$}

\newcommand{\Cpp}{C$^{++}$}

\newcommand{\Np}{N$^{+}$}
\newcommand{\Npp}{N$^{++}$}

\newcommand{\Op}{O$^{+}$}
\newcommand{\Opp}{O$^{++}$}

\newcommand{\Nep}{Ne$^{+}$}
\newcommand{\Nepp}{Ne$^{++}$}
\newcommand{\Sp}{S$^{+}$}
\newcommand{\Spp}{S$^{++}$}

\begin{document}

     \title{Biases in abundance derivations for metal-rich nebulae}

     \author{Gra\.zyna Stasi\'nska\inst{1}}

     \offprints{x}

     \institute{LUTH, Observatoire de Paris-Meudon, 5 Place Jules Jansen, 92195 Meudon, France}
          \date{Received ???; accepted ???}
     \titlerunning{ }
\authorrunning{ }


          \abstract{Using  ab-initio photoionization models of giant \hii\  regions, we test  methods for abundance determinations based on a direct measurement of the electron temperature, now possible even for moderate to high-metallicity objects.  We find that, for metallicities larger than solar, the computed abundances deviate systematically  from the real ones, generally by larger amounts for more metal-rich \hii\ regions. We discuss the reasons for this, and present diagrams allowing the reader to better understand the various factors coming into play. 
We  briefly discuss less classical methods to derive abundances in metal-rich \hii\ regions. In particular, we  comment on the interest of the oxygen and carbon recombination lines. 
We  also show that, contrary to the case of giant \hii\ regions,  the physical conditions in bright extragalactic planetary nebulae are such that their chemical composition can be accurately derived even at high metallicities. Thus, extragalactic planetary nebulae are promising potential probes of the metallicity of the interstellar medium in the internal parts of spiral galaxies as well as in metal-rich elliptical  galaxies.  

}

         \maketitle

\section{Introduction}

Giant \hii\ regions were the first indicators of the presence of abundance gradients across the face of galaxies (Searle 1971, Shields 1974). Many other abundance indicators have been used since then, such as supernova remnants (Dopita et al. 1980, Smith et al 1993), B-supergiants (Monteverde et al. 1997, 2000, Urbaneja et al. 2003), and stellar clusters (Ma et al. 2004, Tiede et al. 2004). A future promising indicator is provided by AGB stars (Cioni \& Habing 2003). However, \hii\ regions remain the most popular tool, both because they are relatively easy to observe and because the interpretation of their spectra in terms of abundances is a priori straightforward. A large number of studies based on giant \hii\ regions has established the existence of abundance gradients in spiral galaxies  (see e.g. Zaritsky, Kennicutt \& Huchra 1994 for a review). The chemical abundances in low metallicity \hii\ regions are generally considered reliable, provided that the spectroscopic data are of good quality, allowing an accurate measurement of the electron temperature \Te\ from forbidden line ratios such as \rOiii. In  cases where a direct determination of the electron temperature is not possible (because the \hii\ regions are faint or  metal-rich or because the spectra are not very deep), the abundances are estimated from statistical methods based on strong lines only.  These statistical methods need to be calibrated (see e.g. Pagel et al. 1979 for an introduction of such methods or Stasi\'nska 2004 for a recent discussion). Many calibrations have been proposed. Discrepancies between abundances derived using the same data but different calibrations can be of a factor around 3 (see e.g. Pindao et al 2002). Calibration is a difficult task which requires a sufficient number of representative \hii\ regions and either detailed modelling (taking into account geometrical effects and accounting for all the relevant data) or a thorough comparison with independently derived, accurate abundances, such as abundances from interstellar absorption lines (Pilyugin 2003). So far, the last word has not been said on these matters.

With the advent of very large telescopes, it is now possible to detect and measure the intensities of very weak lines, including  \Oiiit, \Siiit, or  \Niit\ which are used for electron temperature measurements. It is thus hoped that it will be possible to derive accurate abundances in metal-rich \hii\ regions using classical \Te-based methods. Indeed, a few such measurements have already been published, with the conclusion that the metallicity is not as high as was thought before (Castellanos et al. 2002, Kennicutt et al. 2003, Garnett et al. 2004, Bresolin et al. 2004). 

That the metallicity gradients in the inner parts of disks of galaxies might not be as steep as previously thought is also suggested by other indicators such as infra-red lines in \hii\ regions (Willner \& Nelson-Patel 2002), B stars (Smartt et al., 2001, Munn et al. 2004) and even planetary nebulae (G\'orny et al. 2004). However, the evidence is still based on small data samples, or on methods that are not free from criticism. It is thus important to assess the reliability of the methods used in the case of giant \hii\ regions and to propose alternatives. It has already been shown by Stasi\'nska (1978a, 2002) that abundances derived by \Te-based methods are likely to be biased towards low metallicities when applied to high metallicity \hii\ regions. 

This paper develops on this aspect. For this, we compute series of photoionization models representative of giant \hii\ regions and analyse them with classical \Te-based methods for abundance determinations. We explain the reasons for the bias that occurs for metal-rich \hii\ regions when these methods are applied. We briefly discuss other direct methods of abundance analysis. We next consider extragalactic planetary nebulae as abundance indicators and show that in this case \Te-based abundance determinations are devoid of bias. We finally come back to the case of giant \hii\ regions, and show that their optical spectra are extremely sensitive to local perturbations.  The main points of this study are summarized in Sect. 6.

\section{\Te-based abundance determinations of model giant \hii\ regions}

\subsection{Photoionization models of giant \hii\ regions}

In order to test the reliability of \Te-based abundance determinations, we construct series of photoionisation models in which we vary the metallicity.  The model nebulae are spherical with a central ionizing source. Since our paper is purely methodological it is sufficient to consider ionization  by blackbodies. We consider luminosities that lead to a total number of ionizing photons \qh\ between $4\times 10^{49}$ and $4\times 10^{51}$ s$^{-1}$,  which roughly corresponds to exciting star clusters in giant \hii\ regions. We consider three  geometries: i) a  homogeneous sphere of hydrogen density \nh\ = 50\cmcub; ii) a geometrically thin bubble of density \nh\ = 100\cmcub; iii) a core-halo geometry with an inner region at a constant density \nh\ = 1000\cmcub\ surrounded by a halo where the density decreases as the square of the distance to the ionizing source. 

The elemental abundances are varied in lockstep with those in the Sun as compiled by Lodders (2003), except for nitrogen (we assume that N/O varies proportionally to O/H and has the solar value when O/H is solar) and  helium (assumed constant  and equal to 0.085).  Each series consists of 10 models with oxygen abundances going from log O/H +12 = 8.3 to 9.2 by steps of 0.1~dex. In some series, the abundances of Mg, Si and Fe are multiplied by a factor of 0.01 to simulate the effects of depletion onto dust grains.

The photoionization models are constructed with the code PHOTO as described in  Stasi\'nska \& Leitherer (1996) with atomic data updated as indicated in the Appendix of the present paper. All the models are  ionization bounded (except the core-halo ones)

Table 1 lists the  series of models considered in this work.

\begin{table*}
\caption{Definition of the model sequences
}
\label{models}

\begin{tabular}{llllllllll}

name & \Teff\ & geometry & \nh\ &  \qh\ & $U_{\rm{in}}$ & $U_{\rm{out}}$ &depletion & colour & symbol \\
  & K &   & cm$^{-3}$ & photons s$^{-1}$ &  & & & & \\

\hline
45S1 & 45000 & filled sphere & 50 & $ 4 10^{49}$ & $3.5 10^{1}$ & $3.5 10^{-3}$ &  no & red & squares \\
45S1D & 45000 & filled sphere & 50 & $ 4 10^{49}$ & $3.5 10^{1}$ & $3.5 10^{-3}$ &  0.01 & red & curved squares \\
45S2 & 45000 & filled sphere & 50 & $ 4 10^{51}$ & $1.6 10^{2}$ & $1.6 10^{-2}$ & no & red & circles \\
45B & 45000 & bubble & 100 & $4 10^{51}$  & $ 2.3 10^{-3}$ & $2.2 10^{-3}$ & no & red & triangles \\
45BD & 45000 & bubble & 100 & $4 10^{51}$  & $ 2.3 10^{-3}$ & $2.2 10^{-3}$ & 0.01 & red & diamonds \\
45CH & 45000 & core-halo & 1000-50 & $ 4 10^{51}$ & $4.6 10^{2}$ & $2.0 10^{-1}$ & no & red & stars \\
50S1 & 50000 & filled sphere & 50 & $ 4 10^{49}$ & $4.2 10^{1}$ & $4.2 10^{-3}$ &  no & black & squares \\
50S1D & 50000 & filled sphere & 50 & $ 4 10^{49}$ & $4.2 10^{1}$ & $4.2 10^{-3}$ &  0.01 & black & curved squares \\
50BD & 50000 & bubble & 100 & $4 10^{51}$  & $ 2.7 10^{-3}$ & $2.7 10^{-3}$ & 0.01 & black & diamonds \\
\hline
100PN & 100000 & planetary nebula & 1000 & $ 1 10^{47}$ & $ 1.6 10^{-2}$ & $1.8 10^{-3}$& no & blue & filled circles  \\
150PN & 150000 & planetary nebula & 10000 & $ 5 10^{47}$ & $ 5.5 10^{-2}$ & $4.8 10^{-3}$& no & blue& filled squares \\
\hline
\end{tabular}
\end{table*}

\subsection{The biases}

We  applied to our models the analysis techniques that are used for real objects. We   derived the electron densities from the \rSii\ line ratio and the temperatures \TrOiii, \TrNii\ and \TrOii, respectively, from  the  \rOiii, \rNii\ and \rOiit\ ratios. Then we derived the oxygen abundance using standard \Te-based techniques and assuming that O/H = (\Op\ + \Opp)/\Hp. Note that, as stressed by e.g. Stasi\'nska (1978a),  these temperature estimates  are not identical to the mean ionic temperatures for these ions (defined as \TOpp\
= $\int T_{\rm{e}} n({\rm{O}^{++}}) n_{\rm{e}} {\rm{d}} V   / \int n({\rm{O}^{++}}) n_{\rm{e}} {\rm{d}} V $
for \Opp, and similarly for other ions). The differences may be significant in the case of large temperature gradients or fluctuations. In that case, the temperature characteristic of an emission line, e.g.  \TlOiii\ for the  \Oiii\ line (whose definition is given in Stasi\'nska  1978a)  is different from both the line ratio temperature \TrOiii\ and the mean ionic temperature \TOpp. 

The atomic data used in the abundance derivations are exactly the same as those used in the model computation. This allows us to investigate the biases due to the method, independent of uncertainties in the atomic data (which  are expected to be small, of the order of 10\% or less for the ions we consider).

\begin{figure*}[!hbpt]
\includegraphics[width=8cm]{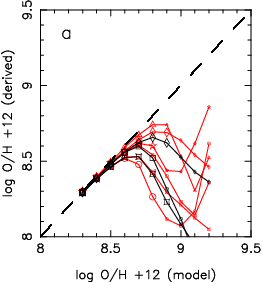}
\includegraphics[width=8cm]{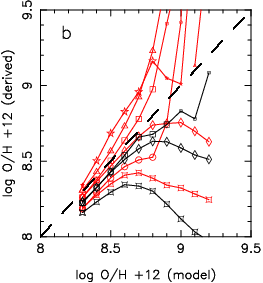}
\caption{The oxygen abundance derived by \Te-based methods for sequences of model \hii\ regions as a function of the oxygen abundance in the models. Each sequence is represented by a different symbol, indicated in Table 1.
In panel a, O/H is  computed using \TrOiii\ and \TrNii. Smaller symbols correspond to models with \Oiiit/\Hb\ $<$  $10^{-4}$. In panel b, O/H is  computed using  \TrNii\ only. Smaller symbols correspond to models with \Niit/\Hb\ $<$  $10^{-4}$. This figure shows that, at high metallicities, the derived O/H values deviate  strongly from the true ones.
}
\end{figure*}

Fig. 1 shows the derived values of log O/H +12 as a function of the input values, for all the models of giant \hii\ regions considered.
In panel a, the abundances were obtained assuming that \TlOiii\ = \TrOiii\ and \TlOii\ = \TrNii, i.e. what an observer would do if he could measure   the \Oiiit\ and \Niit\ line intensities. In panel b, it was assumed that \TlNii\ = \TrNii\ and that \TlOiii\ = (\TlNii\ - 3000) / 0.7, a formula commonly used (e.g.  Campbell et al. 1986, Kennicutt et al. 2003, Garnett et al. 2004a, Bresolin et al. 2004), based on the grid of photoionization models by Stasi\'nska (1982). 
It is seen that, as long as the metallicity is low, the derived O/H value is very close to the input one  (the slight deviations seen in Fig. 2a at low metallicity are due to the fact that the  adopted empirical relation between \TlOiii\ and  \TrNii\ is fulfilled by the models only in first approximation and with some scatter). Important deviations appear around log O/H + 12 = 8.6, and may become huge as the metallicity increases. In the case of Fig. 1a, all the derived values are smaller than the input ones, sometimes by enormous factors. This means that, if such metal-rich \hii\  regions exist, and if observations allow one to measure the corresponding line ratios, the method applied in Fig. 1a will always leads to sub-solar derived oxygen abundances! On the other hand, the method applied in Fig. 1b will either  underestimate or overestimate the oxygen abundance. The bias may exceed 0.2\,dex when the true abundance is larger than the solar one (log O/H + 12 =  8.7, Allende Prieto et al. 2001). In the most extreme cases, however, the transauroral lines are so weak that they might not be observable even with very high signal-to-noise spectroscopy. In Fig. 1a we  represent by smaller symbols the models for which \Oiiit/\Hb\ is smaller than $10^{-4}$, while in Fig. 1b  smaller symbols correspond to models for which \Niit/\Hb\ is smaller  than $10^{-4}$. This is the flux limit  attained by Esteban et al. 2002 with 3 hours of integration on a 4~m telscope for the brightest \hii\ region in M33. We thus see that the best observations available today may lead to considerably biased abundance determinations. 
  In extreme cases, one may use the measured \TrOiii\ value to judge the likeliness of the inferred O/H value. For example, if we find log O/H + 12 = 8.3 with the method used in Fig. 1a while \TrOiii\ is found to be equal to 5000 -- 7000\,K, the true abundance is likely larger than solar, as can be inferred from Fig. 2a, to be described later. However, less extreme cases are more difficult to judge, and would require proper photoionization modelling (but see Sect. 4.4).  

\section{The cause of the abundance biases at high metallicities}

\subsection{Temperature gradients inside the nebulae}

It has long been known (Stasi\'nska 1978b, 1980, Garnett 1992) that at high metallicities one expects large temperature gradients inside \hii\ regions, due to the fact that the inner zone contains more efficient coolants (principally \Opp\ through its infra-red lines) than the outer zones. 
In the following, we elaborate on the causes and consequences of these  temperature gradients.

\begin{figure*}[!hbpt]
\includegraphics[width=8cm]{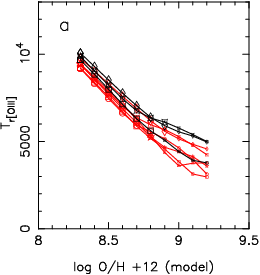}
\includegraphics[width=8cm]{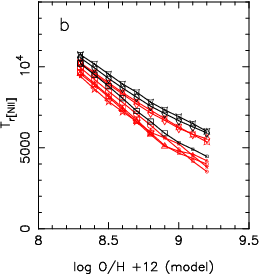}
\caption{Panel a: \TrOiii\ as a function of O/H for our model \hii\ regions. Panel b: \TrNii\ as a function of O/H. Same symbols as in Fig. 1. }
\end{figure*}

\begin{figure*}[!hbpt]
\includegraphics[width=16cm]{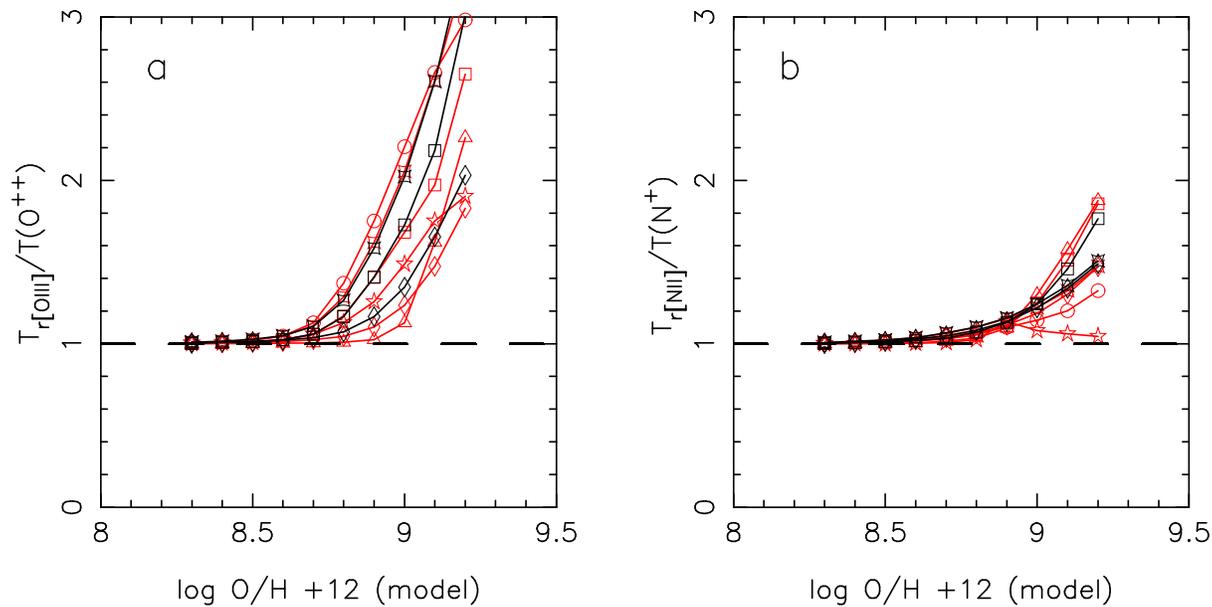}
\caption{Panel a: \TrOiii/$T$(\Opp) as a function of O/H for our model \hii\ regions. Panel b: \TrNii/$T$(\Np) as a function of O/H. Same symbols as in Fig. 1. This figure illustrates that, at high metallicity,  \TrOiii\ and  \TrNii\ are not representative of the \Opp\ and \Np\ zones, respectively.}
\end{figure*}

Fig. 2a shows \TrOiii\ as a function of log O/H +12 while  Fig. 2b shows \TrNii\ as a function of log O/H +12. We see that both temperatures decrease with increasing abundances. This is due to the enhanced cooling by heavy elements.  At high metallicities, however, neither \TrOiii\ nor \TrNii\ are close to the corresponding mean ionic temperatures, as seen in Figs. 3a and 3b which show \TrOiii/\TOpp\ and \TrNii/\TNp, respectively, as a function of log O/H +12. The values of \TrNii\ and \TrOiii\ strongly overestimate the values of \TNp\ and \TOpp\ at metallicities above log O/H +12 =8.8. The models show that they also significantly overestimate the values of \TlNii\ and \TlOiii.

\begin{figure*}[!hbpt]
\includegraphics[width=8cm]{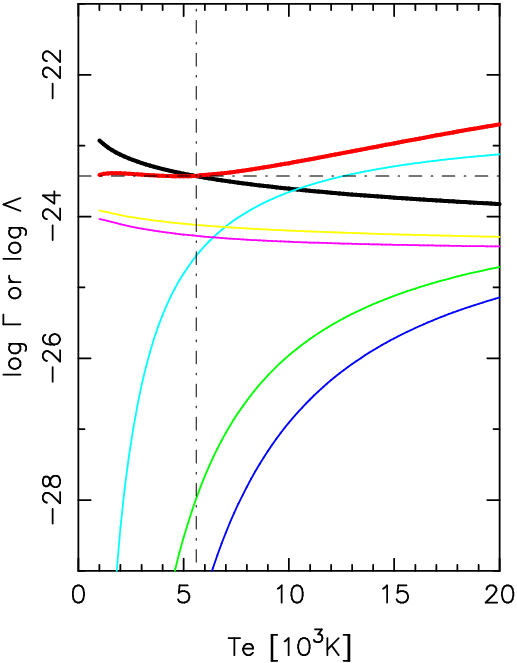}
\includegraphics[width=8cm]{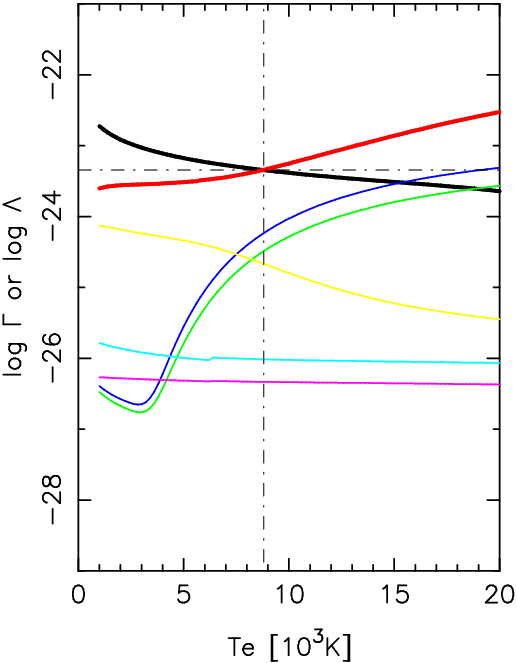}
\caption{Energy losses (thick increasing curve) and energy gains (thick decreasing curve) as a function of \Te, in the 45S2 \hii\ region model at solar metallicity.  Panel a:   high ionization zone (\Opp/\Op\ = 0.9). The thin curves represent the emissivities  O~{\sc iii}] $\lambda$1661 (green),   [O~{\sc iii}] $\lambda$4363 (blue),  [O~{\sc iii}] $\lambda$4949, 5007 (cyan),   [O\,{\sc{iii}}]\,$\lambda$52\,$\mu$m (magenta),  [O\,{\sc{iii}}]\,$\lambda$88\,$\mu$m (yellow).  Panel b:  low ionization zone (\Opp/\Op\ = 0.1). The thin curves represent the emissivities of [O~{\sc ii}] $\lambda$3729 (green),  [O~{\sc ii}] $\lambda$3726 (blue),  [N~{\sc ii}] $\lambda$121\,$\mu$m (cyan),  [N\,{\sc{ii}}]\,$\lambda$205\,$\mu$m (magenta),  [Si\,{\sc{ii}}]\,$\lambda$35\,$\mu$m (yellow).}
\end{figure*}

It is useful to visualize why temperature gradients are predicted to be so large in high metallicity \hii\ regions. Let us take as an example model 45S2 at solar metallicity, and consider one point in the high excitation region defined by a local \Opp/\Op\ ratio of 0.9, and one point in the low excitation region, defined by a local \Opp/\Op\ ratio of 0.1. We plot in Fig. 4a the \Te\ dependence of log $\Gamma$ and of log $\Lambda$ in the high excitation region ($\Gamma$ and $\Lambda$ are, respectively, the total energy gains and losses by the  gas, per unit time, per electron and per hydrogen particle). Both $\Gamma$ and $\Lambda$ are in ergs cm$^3$ s$^{-1}$. Fig. 4b shows the same but for the low excitation zone, defined by a local \Opp/\Op\ ratio of 0.1. In Fig. 4a, we also plot the emission in the \Opp\ lines  per unit time, per electron and per hydrogen atom. The various lines are represented in different colours (visible in the on-line version), with the following coding: green: O~{\sc iii}] $\lambda$1661, blue: [O~{\sc iii}] $\lambda$4363,  cyan: [O~{\sc iii}] $\lambda$4949, 5007,  magenta: [O\,{\sc{iii}}]\,$\lambda$52\,$\mu$m, yellow: [O\,{\sc{iii}}]\,$\lambda$88\,$\mu$m. In this figure, all the curves include the change in the ionization balance due to the variation in electron temperature. We see that the oxygen lines are a major contributor to the cooling in the high excitation region. The electron temperature at each point in a nebula is defined by the condition $\Gamma$ = $\Lambda$. It can  easily be shown (see Osterbrock 1989 or Stasi\'nska 2002) that $\Gamma$ is roughly proportional to the effective temperature of the exciting stars and to the recombination coefficient of hydrogen, which varies approximately like \Te$^{-1}$. The temperature dependence of  $\Lambda$ results from the sum of the contribution of the different cooling lines. At high temperatures, the losses are dominated by \Oiii\ and increase with \Te, while at low temperatures they are dominated by the infrared oxygen lines, and are almost independent of \Te. In the low excitation region, oxygen is less dominant in the loss processes. The thin curves in Fig. 4b represent  the emission in the most important cooling lines, with the following colour coding: 
green: [O~{\sc ii}] $\lambda$3729, blue: [O~{\sc ii}] $\lambda$3726,  cyan: [N~{\sc ii}] $\lambda$121\,$\mu$m,  magenta: [N\,{\sc{ii}}]\,$\lambda$205\,$\mu$m, yellow: [Si\,{\sc{ii}}]\,$\lambda$35\,$\mu$m. Note that the temperature in the low ionization zone,  8700\,K, is higher than in the high ionization zone,  5500\,K,  both because cooling is less efficient and  because heating is more efficient due to the hardening of the radiation field in the outskirts of the nebula. At low temperatures, [Si\,{\sc{ii}}]\,$\lambda$35\,$\mu$m heavily contributes to the losses in the low ionization zone, which implies that the temperature of the gas is strongly dependent on the degree of depletion of Si onto grains (as already pointed out by Henry 1993 and Shields \& Kennicutt 1995).  For a higher metallicity than represented in Fig. 4, the entire loss curves will be shifted upwards by a quantity roughly equal to the log of the metallicity increase, and the electron temperature, defined by the condition $\Gamma$ = $\Lambda$, will decrease. For an increase in metallicity by a factor of two from the solar value,  the temperature will drop from \Te\ = 5500\,K to \Te\ = 1500\,K in the high ionization zone, and from \Te\ = 8700\,K to \Te\ = 5000\,K in the low ionization zone, resulting in a  larger \Te\ gradient than in the case of solar metallicity. Note that, at the equilibrium temperature  for twice solar metallicity, the optical and ultraviolet lines from \Opp\ ions are practically not emitted in the high ionization region. Most of the emission in these lines  actually comes from a layer containing mostly \Op, so that the temperature is sufficiently large to allow collisional excitation of these lines. Therefore \TrOiii\ and \TlOiii\ are not representative of the \Opp\ zone.

Fig. 4 also allows one to understand why, in the case of large temperature gradients, \TrOiii\ is larger than \TlOiii, the temperature characteristic of the \Oiii\ line: the \rOiii\ ratio is largely weighted by the zones of highest temperatures.

The expected low temperatures and large temperature gradients at high metallicities have dramatic consequences for abundance determinations from optical lines. The derived  abundances can be under- or overestimated according to whether the temperatures adopted to derive the line emissivities actually over- or underestimate the true line temperatures. The second case may occur if the temperature in the low excitation zone is estimated not directly from observations but from inappropriate model grids, as is the case in Fig. 1b. When both \TrOiii\ and \TrNii\ can be measured, the \Te-method will strongly underestimate the abundances (as seen in Fig. 1a). 

\subsection{Contribution of recombination to collisionally excited lines.}

\begin{figure}[!hbpt]
\includegraphics[width=8cm]{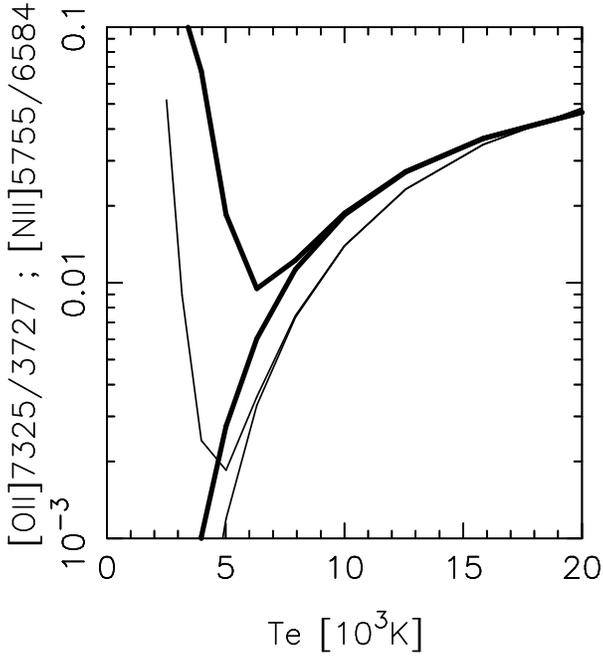}
\caption{The \Te\ dependence of the  \rNii\ ratio (thin curve) and of the  \rOiit\ ratio (thick curve). The monotonic curves ignore the contribution of recombination. The non-monotonic curves have been computed taking into account recombination from \Npp\ and \Opp\ and assuming that these ions are present in the same proportion as \Np\ and \Op.}
\end{figure}

Another factor that affects abundance determinations is the contribution of recombination to the emissivity of forbidden lines. This has been shown to be important in planetary nebulae with cool, hydrogen-poor inclusions (Liu et al. 2000, Tsamis et al. 2003a). The importance of this effect in metal-rich \hii\ regions has not been fully recognized, although it has been considered by Kennicutt et al. 2003. These authors have corrected the \Oiitoneb\ line emissivity for recombination using the formula given by Liu et al. (2000): \Oiitoneb/\Hb\ = 9.36 (\Te/$10^4$)$^{0.44}$ \Opp/\Hp\ (which nb. is valid only for \Te/$10^4$ between 0.5 and 1).  For high metallicity \hii\ regions this formula may lead to considerable errors. First, if \Opp/\Hp\ is obtained from collisionally excited lines, it will be overestimated. Second, the electron temperature characteristic of the \Hb\ emission is different from the one characteristic of the \Oiitoneb\ emission.

In Fig. 5 we illustrate the effect of recombinations on the temperatures derived from \rNii\ and \rOiit. The thin curves represent the \rNii\ ratio  and the thick ones the  \rOiit\ ratio. The non-monotonic curves have been computed taking into account recombination from \Npp\ and \Opp\ and assuming that these ions are present in the same proportion as \Np\ and \Op. The monotonic curves have been computed ignoring recombination (these are the ones generally used for temperature determinations).  We see that both ratios can be strongly affected by recombination. Below a certain temperature, these ratios are  even dominated by recombination.  The minimum of the   \rOiit\ ratio occurs at an electron temperature of around 6000\,K, which can be obtained in the \Op\ Êzones for metallicities of the order of  log O/H + 12 = 8.75 -- 8.9. For \rNii, the minimum occurs at 5000\,K, corresponding to O/H + 12 of  8.9 -- 9. 

Note that forbidden line from \Sp\ and \Spp\ are also expected to be affected by recombination. However, atomic data that would make it possible to compute the contribution of recombination are not yet available for these ions.

\section{How to eliminate the biases in abundance determinations}

\subsection{Derive abundances from far infrared lines}

Infrared line emissivities of metals are almost independent of temperature, and thus in principle more suitable than optical lines to derive abundances in the case of metal-rich \hii\ regions. Their interpretation poses other problems, though (Simpson et al. 2004). So far, studies of extragalactic \hii\ regions using far infrared lines are scarce.  Willner \& Nelson-Patel (2002) have used the measurements of \neii\ and \neiii\ infrared lines in 25 \hii\ regions in M33 to derive the neon abundance gradient. They found that this gradient is much shallower than the oxygen abundance gradient derived  from optical data. However, their  Ne/H ratios were obtained assuming a temperature of $10^{4}$\,K. This assumption creates a bias, since high metallicity \hii\ regions will have smaller \Te\ than low metallicity ones.  In the case when the average electron temperature is 5000\,K, which corresponds to our \hii\ region models having log O/H +12 between 8.7 and 8.9, the procedure adopted by Willner \& Nelson-Patel underestimates the neon abundance by a factor 2. If the average temperature of the \hii\ gas could be measured accurately then empirical methods using infrared lines would provide accurate neon abundances in \hii\ regions.

\subsection{Measure the average electron temperature of the entire nebula}

\begin{figure}[!hbpt]
\includegraphics[width=8cm]{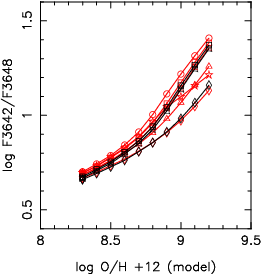}
\caption{The Balmer jump as a function of O/H for our model \hii\ regions.}
\end{figure}

The Balmer jump is sensitive to electron temperature and is very large at low electron temperatures. Therefore, in principle, it should be a good temperature indicator at high metallicities. Fig. 6 shows the value of the Balmer jump, defined as the ratio of nebular fluxes at 3642 and 3648~\AA, for our various sequences of models. The measurement of the Balmer jump  in supposedly metal-rich extragalactic \hii\ regions should be an important element in deriving accurate abundances. Unfortunately, this measurement is difficult, and the contribution of the stellar populations must be adequately subtracted.

\subsection{Derive abundances from pure recombination lines}

\begin{figure*}[!hbpt]
\includegraphics[width=16cm]{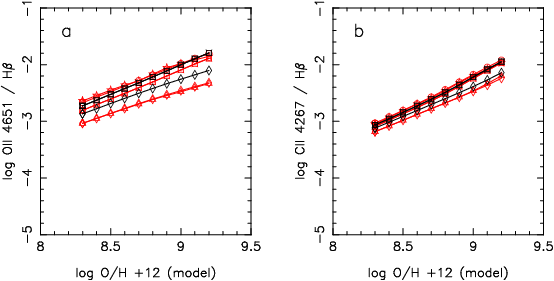}
\includegraphics[width=16cm]{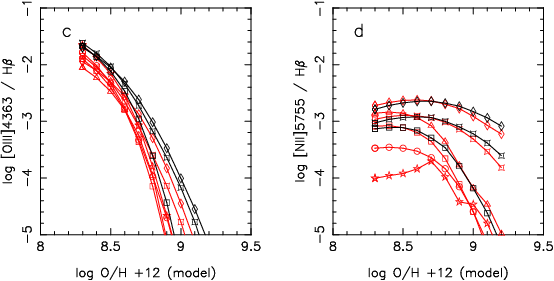}
\caption{The intensities of some weak lines as a function of O/H for our model \hii\ regions. Panel a:  O~{\sc ii} $\lambda$4651/\Hb. Panel b: C~{\sc ii} $\lambda$4267/\Hb. Panel c: \Oiiit/\Hb. Panel d: \Niit/\Hb. }
\end{figure*}

At metallicities above solar, our models of giant \hii\ regions predict for the recombination lines C~{\sc ii} $\lambda$4267 and O~{\sc ii} $\lambda$4651 intensities of up to one hundredth of \Hb. This is  much stronger than the predicted intensities for the \Oiiit\ or \Niit\ lines, as seen in Fig. 7. The intensity ratios of recombination lines have the virtue of being only little dependent on the electron temperature, therefore abundances obtained from recombination lines are considered accurate.  Some bias may exist, however, in the case of nebulae with important temperature stratification, since the \Cpp\ and \Opp\ ions, which generate these two lines, are present only in the high ionization zones while \Hb\ is emitted everywhere. 

It must be noted, however, that the interpretation of recombination lines of heavy elements involves other problems. In planetary nebulae,  abundances derived from recombination lines  are systematically larger than abundances derived from collisionally excited lines by factors from 1.5 to 10 (see e.g. Liu 2002 for a review). Recombination lines have also been detected in a few \hii\ regions (e.g. Peimbert et al. 1993, Esteban et al. 1998, 1999, 2002, Tsamis et al. 2003b, Garcia-Rojas et al. 2004). Here again they lead to higher abundances than collisionally excited lines, although not to the same extreme as in planetary nebulae. Temperature stratification cannot be the explanation, since this discrepancy has also been found in Magellanic Clouds \hii\ regions that are not expected to harbour strong temperature gradients  because of their subsolar metallicities. Tsamis et al. (2003b) make the point that  this discrepancy cannot be attributed to temperature fluctuations such as described by Peimbert (1967). Presently, the most favoured explanation is chemical composition inhomogeneities (e.g. Liu et al. 2000, Tsamis et al. 2003b) involving extremely metal-rich clumps, whose nature is not necessarily the same in planetary nebulae and in \hii\ regions. 

Therefore  recombination lines, while offering important additional clues to the heavy element abundances in \hii\ regions, are not easy to interpret.

\subsection{Tailored photoionization modelling}

In principle, photoionization models take into account all the processes that affect the thermal balance at each point in the nebula. They also take into account recombination for the emission of C, N, O and Ne, at least PHOTO and CLOUDY  (Ferland  1996) do. Therefore, if one fits the observed emission line ratios with a tailored model, the concerns expressed above are not relevant. To simplify the problem, let us assume that only oxygen plays an important role in the thermal balance. The main parameters defining a model are the mean effective temperature of the stars, the  characteristic ionization parameter, and the oxygen abundance, i.e. 3 parameters. The main observational constraints are \Oiii/\Hb, \Oiiit/\Hb, \Oii/\Hb\ and \Oiit/\Hb, i.e.  in principle more than enough in this simple representation of \hii\ regions (since the measurement of \Oiiit\ Êalleviates the double-value problem encountered by strong line methods). 

Castellanos et al. (2002) have produced simple tailored models for a dozen supposedly metal-rich giant \hii\ regions assuming a simple bubble geometry, and a reasonable solution could be found in many cases (although the optimum number of constraints was not always available), with the result that the metallicity appears to be below solar in the majority. Actually, even without   detailed photoionization modelling, the fact that most of the objects studied by Castellanos et al. (2002) have \TrOiii\ and/or \TrNii\ larger than 9000\,K implies that their metallicities are not above solar. This can be seen directly on our Figs. 2a and 2b (the use of a more realistic description of the stellar radiation field in these figures would not affect this claim severely ). Of course, an accurate abundance derivation requires a correct model-fitting.

However, the fact that models are not always able to reproduce the optical spectra even in  low metallicity giant \hii\ regions  (e.g. Luridiana et al. 1999, Stasi\'nska \& Schaerer 1999) should warn us against being too confident in models for high metallicity \hii\ regions. For the moment,  the thermal balance of \hii\ regions is not fully understood, and this lack of understanding might have dramatic consequences for the determination of abundances in metal-rich \hii\ regions, where the optical forbidden line emissivities depend so strongly on electron temperature.

Other factors, such as the density distribution of the nebular gas and especially the degree of depletion of metals onto grains and the dust content (Henry 1993, Shields \& Kennicutt 1995), also affect the spectrum of \hii\ regions and call for  a larger number of observational constraints   (e.g. mid- and far-infrared data). 

If indeed large temperature gradients in metal rich \hii\ regions exist, the  spatial distribution of emission line ratios would be a useful constraint, as shown by Stasi\'nska (1980). Because of the extremely low temperatures in the high ionization zones, collisionally excited lines  are expected to be strong only in low ionization zones (see Fig. 4). Therefore, contrary to what is observed at subsolar metallicities,   the \Oiii\ and \Hei\ emitting zones should not be  coextensive. Note that this property does not depend on the details of the nebular geometry.

Another important constraint on the models comes from infrared lines. Thanks to ISO and the Spitzer Space Telescope, these are becoming available for extragalactic \hii.  So far, the only tailored modelling of an extragalactic \hii\ region that uses optical spectra and images as well as infrared line intensities are the ones developed by  Garnett et al. (2004b) for a supposedly metal-rich giant \hii\ region in M 51 and by Jamet et al. (2004) for a metal-poor giant \hii\ region in M33. In the first case, the optical data alone did not provide any direct  information on the electron temperature.  Constructing  photoionization models, the authors estimated the oxygen abundance that was able to reproduce the observed \Oiii/\Oiiitoneb\ ratio. However,  the observational constraints for this \hii\ region were not many and the gas density distribution was not explored.  On the other hand, Jamet et al. (2004) studied an \hii\ region with measured \Oiiit/\Oiii\ ratio, and constructed photoionization models constrained by the observed population of the exciting stars and by the observed nebular geometry. They were not able to reproduce concomitantly the optical and infrared line fluxes, implying that the uncertainty in abundances is larger than thought, even in the case of low metallicities.

\section{The use of planetary nebulae as substitutes for \hii\ regions}

As we have seen above, the reason why, in truly metal-rich \hii\ regions, abundances derived using empirical \Te-based  methods are likely to be  biased is due to the very low electron temperature expected in a large fraction of the volume of the \hii\ region. At such a low temperature, collisional excitation of optical lines does not occur, and the integrated emission of \oiii\ lines is strongly weighted by the hottest zones of the \hii\ region, resulting in the  \oiii\ lines being actually emitted where \Opp\ is present only in small amounts.

Such a low temperature in the \Opp\ zone of metal-rich nebulae would not occur if the nebulae were ionized by much hotter stars (increased heating) and if they were denser (reduced collisional deexcitation of infrared cooling lines). 
Extragalactic planetary nebulae that are bright enough for high signal-to-noise spectroscopy present both these advantages. It can be shown, by using simple photoionization models for expanding planetary nebulae surrounding evolving post-AGB stars (Stasi\'nska et al. 2004) that the most luminous planetary nebulae (i.e. the ones for which the largest number of important diagnostic lines can be observed) are nebulae that have central stars with temperatures above 100000\,K, and hydrogen densities above 1000\cmcub. 

We have constructed two series of models of planetary nebulae fulfilling these requirements. They are presented at the bottom of Table 1. The set of elemental abundances is identical to the one adopted for the giant \hii\ regions models. 

\begin{figure*}[!hbpt]
\includegraphics[width=8cm]{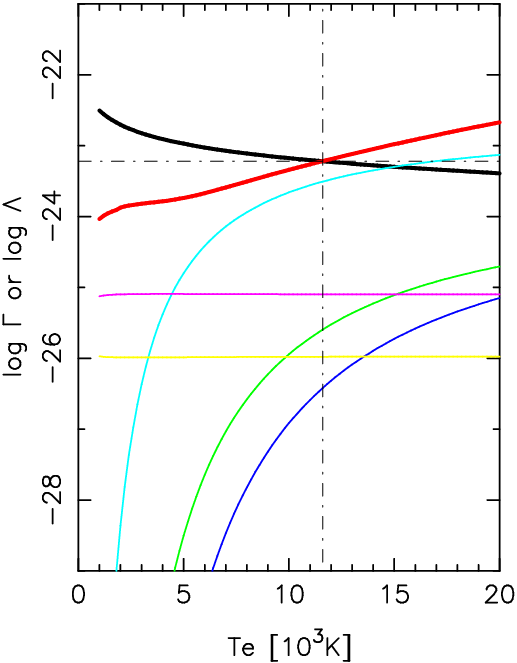}
\includegraphics[width=8cm]{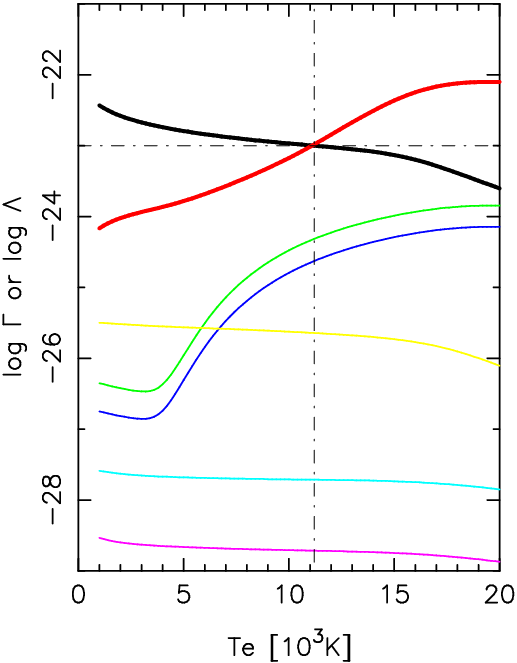}
\caption{
Energy losses (thick increasing curve) and energy gains (thick decreasing curve) as a function of \Te, in the 150PN planetary nebula   model at solar metallicity.  Panel a:   high ionization zone (\Opp/\Op\ = 0.9). The thin curves represent the emissivities  O~{\sc iii}] $\lambda$1661 (green),   [O~{\sc iii}] $\lambda$4363 (blue),  [O~{\sc iii}] $\lambda$4949, 5007 (cyan),   [O\,{\sc{iii}}]\,$\lambda$52\,$\mu$m (magenta),  [O\,{\sc{iii}}]\,$\lambda$88\,$\mu$m (yellow).  Panel b:  low ionization zone (\Opp/\Op\ = 0.1). The thin curves represent the emissivities of [O~{\sc ii}] $\lambda$3729 (green),  [O~{\sc ii}] $\lambda$3726 (blue),  [N~{\sc ii}] $\lambda$121\,$\mu$m (cyan),  [N\,{\sc{ii}}]\,$\lambda$205\,$\mu$m (magenta),  [Si\,{\sc{ii}}]\,$\lambda$35\,$\mu$m (yellow).}
\end{figure*}

Figs. 8a and 8b show the energy gains and losses in the same way as Fig. 4a and 4b but for the solar abundance planetary nebula model in the 150PN series instead of the model \hii\ region. These figures show that the temperatures in the \Opp\ and \Op\ zones are similar, and that the \Oiii\ and \Oiiit\ lines are easily emitted in the  \Opp\ zone, so that \TrOiii\ is really representative of the temperature in the \Opp\ zone (and also in the \Op\ zone to a good approximation). Comparing with Fig. 4, $\Gamma$ has increased by about a factor 3 and $\Lambda$ has decreased at the low temperature end. These changes lead to an equilibrium temperature of 11500\,K (instead of 5500\,K  for the giant \hii\ region model at same metallicity shown in Fig. 4). At 11500\,K, the \Oiii\ and \Oiiit\ are emitted  efficiently. 

\begin{figure*}[!hbpt]
\includegraphics[width=8cm]{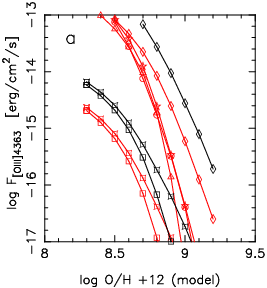}
\includegraphics[width=8cm]{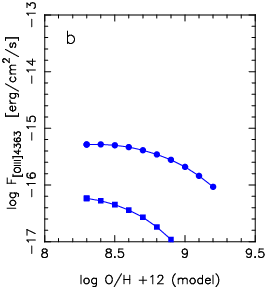}
\caption{The total flux in the \Oiiit\ line at a distance of 1\,Mpc. Panel a: for our model \hii\ regions. Panel b: for our model planetary nebulae.}
\end{figure*}

Despite the orders of magnitude difference between intrinsic luminosites of planetary nebulae and giant \hii\ regions,  the \Oiiit\ line is actually stronger in bright planetary nebulae than in giant \hii\ regions of the same metallicity when metallicities are larger than solar. This is shown in Fig. 9, where we plot the \Oiiit\ fluxes from our model giant \hii\ regions (panel a) and planetary nebulae (panel b), supposed located at a distance of 1~Mpc. 

\begin{figure}[!hbpt]
\includegraphics[width=8cm]{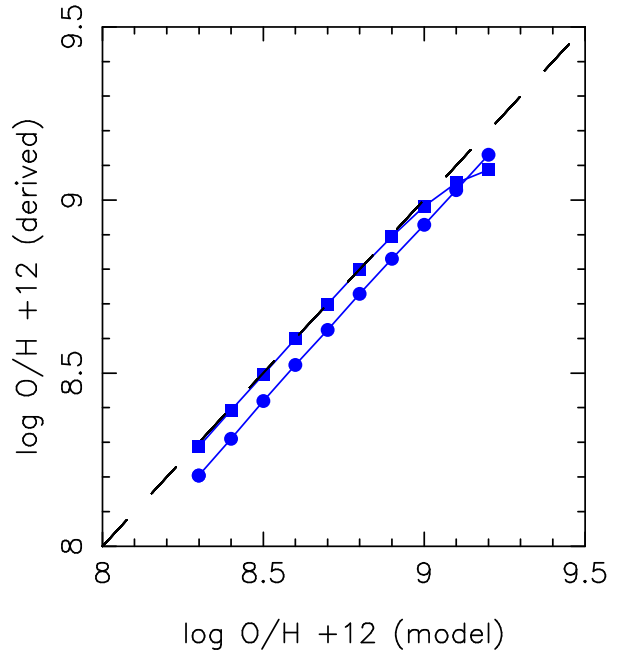}
\caption{The oxygen abundance derived by \Te-based methods for our model planetary nebulae as a function of the oxygen abundance with which the models have been computed. The abundances are computed using \TrOiii\ and \TrNii.}
\end{figure}

Similarly to what we did with our \hii\ region models, we derived the abundances from our planetary nebula models using the classical \Te-based method. Here, we assumed that  \TlOiii\ and  \TlOii\ are equal to \TrOiii. For oxygen, we employed the ionization correction factor commonly used for planetary nebulae, i.e. : O/H = (\Op\ + \Opp)/\Hp\ $\times$ [(\Hep\ + \Hepp)/\Hep]$^{2/3}$ (Kingsburgh \& Barlow  1994).
Fig 10 shows the derived values of log O/H +12 as a function of the input values, for both our series of models of planetary nebulae. 
It is seen that the derived values of log O/H +12 are very close to the input ones in the entire metallicity range. For the model with higher \Teff\ and higer ionization parameter (model 150PN in Table 1), the derived values are systematically lower than the input ones by 0.1 dex, because of the inadequacy of the adopted ionization correction factor for this precise model. This error is however much smaller than the errors in derived abundances for giant \hii\ regions, and occurs only at very high effective temperatures and ionization parameters.

The reason why abundances  derived in planetary nebulae are so  accurate  compared to abundances derived in giant \hii\ regions is the larger electron temperature and the smaller internal temperature gradients. Note that, given the large values of \Te\ even for the highest metallicity models, the \Oii\ line is not strongly affected by recombination, contrary to what occurs in metal-rich \hii\ regions.

\begin{figure}[!hbpt]
\includegraphics[width=8cm]{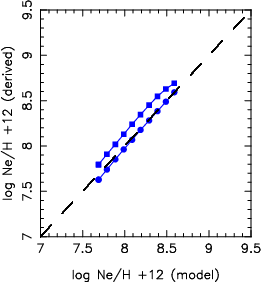}
\caption{The Ne/H ratio derived by \Te-based methods for our model planetary nebulae as a function of input Ne/H values in our models. }
\end{figure}

There is some concern that, in planetary nebulae, oxygen might not trace the abundance of the interstellar medium exactly, principally due to nuclear and mixing processes in the progenitor star (Dopita et al. 1997). If this were true, planetary nebulae would not be convenient objects for studying the metallicity gradients in galaxies. On the other hand, there is no such problem with neon (Marigo et al. 2003). It is therefore interesting to derive  the neon abundance for our model planetary nebulae. Fig. 11 shows the derived values of log Ne/H + 12 as a function of the input ones, assuming the commonly used relation Ne/O = \Nepp/\Opp. We see that the derived Ne/H values are within 0.15\,dex of the input ones, and that no systematic deviation appears with metallicity. For comparison, we show in Fig. 12a and b the derived values of log Ne/H +12 for our model \hii\ regions adopting the same laws for the temperature as for O/H in Figs 1a and b, respectively. We see that Ne/H deviations from the input values are smaller than in the case of O/H, but that they can still be significant. At the highest metallicities, there is a steep rise in the derived Ne/H, even if O/H is underestimated (compare Figs 1a,b and Figs.12a,b). This is due to photoionization of \Nep\ to excited levels of \Nepp, which dominates the production of the  \Neiii\ line at temperatures below, say 3000 -- 4000\,K  (of course, in such cases, the \Neiii\ line is quite weak, of the order of $10^{-3}$ -- $10^{-2}$ of \Hb).
 
\begin{figure*}[!hbpt]
\includegraphics[width=8cm]{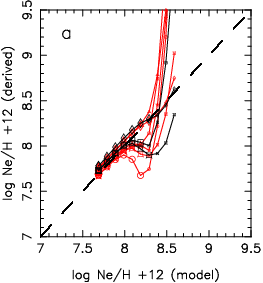}
\includegraphics[width=8cm]{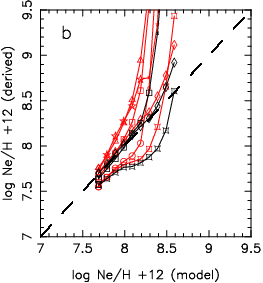}
\caption{The Ne/H ratio derived by \Te-based methods for our model \hii\ regions as a function of input Ne/H values in our models. In panel a, the abundances are computed using \TrOiii\ and \TrNii. Smaller symbols correspond to models with \Oiiit/\Hb\ $<$  $10^{-4}$. In panel b, they  are computed using  \TrNii\ only. Smaller symbols correspond to models with \Niit/\Hb\ $<$  $10^{-4}$.}
\end{figure*}

\subsection{A planetary nebula in a giant \hii\ region}

Real giant \hii\ regions are obviously more complex than the idealized models we presented here. First of all, their geometries, as can be judged from high resolution images (Walborn et al. 2002, Maiz-Apellaniz et al. 2004) can be quite intricate. Besides, the region covered by the spectral slit is likely to contain
emission zones that are not only affected by the ionizing photons from massive stars, but may contain supernova remnants or planetary nebulae. These are not necessarily detected individually, but they may have an effect on the observed giant \hii\ region spectrum. 

\begin{figure*}[!hbpt]
\includegraphics[width=16cm]{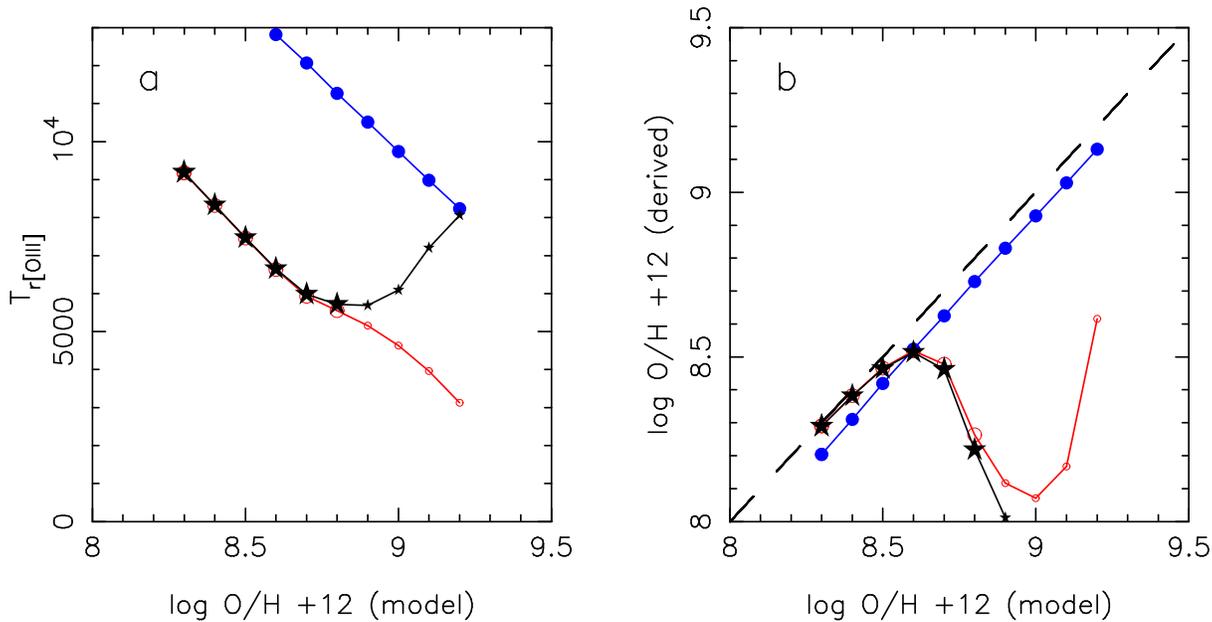}
\caption{ The giant \hii\ region 45S2 alone (red circles), the planetary nebula 150PN alone (filled blue circles), and the sum of both (black stars). Panel a: \TrOiii\ as a function of O/H. Panel b: the derived value of O/H as a function of the O/H value in the model.}
\end{figure*}

Let us consider again Fig. 4a, to understand qualitatively the effect that an intervening planetary nebula may have on the spectrum of an \hii\ Êregion. In an \hii\ region of half solar metallicity, the energy loss curve will drop by about 0.2 dex below the one represented in Fig. 4a (which corresponds to solar metallicity).  In the planetary nebula, where the heating is larger by a factor about three \Te\ will rise to about 13000\,K (assuming that the ionization stage is unchanged).  The \Oiii\ emissivity will thus rise by about a factor of two and the \Oiiit\ emissivity by about a factor of 8. On the other hand, in an \hii\ region of twice solar metallicity, the same local rise in the energy gains will lead to a local increase in the electron temperature from about 2000\,K to about 5000\,K. This implies an increase in the \Oiii\ emissivity by over 4 orders of magnitude  and in the \Oiiit\ emissivity by an even larger amount! Thus, at high metallicities,  optical spectra of giant \hii\ regions are extremely sensitive to local variations of the heating rate. This is likely to induce important biases when deriving their chemical composition. 

We illustrate this by considering the superimposition of  a giant \hii\ region represented by our model 45S2 and of a planetary nebula represented by our model 150PN. We report in Fig. 13a the values of \TrOiii\ as a function of log O/H + 12 computed for the giant \hii\ region alone (red circles), for the planetary nebula alone (filled blue circles), and for the sum of both (black stars).  We see that, at low metallicity, the value of \TrOiii\ is not affected by the presence of the planetary nebula. When log O/H +12 exceeds  8.7, \TrOiii\ starts increasing with O/H. At log O/H +12 = 9.2, it reaches the value corresponding to the planetary nebula alone. As a consequence, at a metallicity larger than solar, the bias on O/H would be  even more dramatic than in the case of a pure giant HII region. This assumes that one is able to measure the \Oiiit\ line intensity. As in Fig. 1a, smaller symbols correspond to the cases where \Oiii/\Hb\ is smaller than $10^{-4}$. We conclude that this ''enhanced bias'' threatens abundance determinations based on the highest quality \hii\ region spectra obtainable today.

\section{Discussion}

Present-day observations allow one to measure line intensities in extragalactic giant \hii\ regions  down to $10^{-4}$ of \Hb. This makes it possible to apply \Te-based methods to derive  abundances even in metal-rich \hii\ regions, while in the past only  strong line methods could be used. Strong line methods are not entirely satisfactory. First of all, they are statistical: they assume that all the \hii\ Êregions obey the same empirical relation between the ionizing radiation field, the nebular geometry  and its metallicity, which may be true only to a certain extent. Besides, the different calibrations proposed in the past yield significantly different results for the derived abundances, sometimes by factors of 3 or more. This is why the possibility of using direct, \Te-based methods -- believed more precise and more accurate -- has raised considerable interest. 

In this paper we have examined the biases involved in \Te-based abundance determinations at high metallicities. Using  photoionization models of giant \hii\ regions, we have tested these methods by applying to the models the same methods as used for real objects. We have found that, for log O/H+12 larger than 8.7 (i.e. larger than the solar value), the computed abundances may deviate from the real ones by factors exceeding 0.2\,dex.  If both \TrOiii\ and \TrNii\ values are used for the abundance determinations,  the computed O/H values are below the true ones, because of strong temperature stratification occurring in our models. If not all the \Te\ diagnostics are used but guesses are made  for some ions, the derived abundances may be either underestimated or overestimated, again by significant amounts. We have discussed the cause for this, and presented diagrams allowing the reader to visualize the various factors coming into play. The bottom line of the reason for the failure of \Te-based methods in the case of metal-rich \hii\ regions is the fact that optical forbidden lines are significantly emitted only above a temperature threshold, and that this threshold lies among the temperatures expected in metal-rich \hii\ regions.

All our analysis is based on ab initio photoionization models that have a simple geometrical structure. In real \hii\ regions, the relative importance of the various factors determining the electron temperature in the different zones of the nebulae may not be exactly the same. One important problem, not considered here, is the effect of dust grains on the thermal balance. 

Because of the temperature dependance of optical forbidden lines and the range of expected values of \Te\ in high metallicity giant \hii\ regions, any local perturbation of the heating rate (shocks, conductive heating) is likely to  affect \Te-based abundances more strongly than in the case of metal-poor \hii\ regions. We have shown that even an intervening high metallicity planetary nebula may affect \Te-based abundance diagnostics of high metallicity giant \hii\ regions.

We have briefly discussed less classical methods to derive abundances in metal-rich \hii\ regions. In particular, we have commented on the interest of the oxygen and carbon recombination lines   which should be more easily measured in metal-rich \hii\ regions than the weak forbidden lines used to estimate electron temperatures.

Detailed photoionization model fitting of presumed metal-rich \hii\ regions for which a large number of observational constraints can be gathered would be important to better test our understanding of the thermal balance of such objects and the reliablility of empirical abundance determinations in them. We note that, for the moment, even low metallicity giant \hii\ regions are not fully understood from this point of view.

  From the astrophyscal point of view, the main impact of abundance determinations in metal-rich \hii\ regions is on the determination of abundance gradients in spiral galaxies. While we have shown that strong biases are likely to occur, we have not discussed the statistical distribution of these biases. As a matter of fact,  some insight into the true abundances might be gained by considering not only the measured abundance gradients (or line intensity gradients) but also their dispersion as a function of galactocentric distance. Such an approach requires large samples of \hii\ regions with adequate measurements, like those secured recently by Cedr{\' e}s \& Cepa (2002) for the galaxies NGC 5457 and NGC 4395. It also requires an a priori knowledge of the estimated abundance dispersion in the interstellar medium of galaxies at a given galactocentric distance, derived from models of chemical evolution of galaxies. 

We have also shown that, contrary to the case of giant \hii\ regions,  the physical conditions in bright extragalactic planetary nebulae are such that their chemical composition can be accurately derived even at high metallicities. Thus, extragalactic planetary nebulae should be investigated as possible probes of the metallicity of the interstellar medium in the internal parts of spiral galaxies as well as in metal-rich elliptical  galaxies.

\begin{acknowledgements}

Part of this work was done while the author was participating in the Guillermo Haro workshop 2004 at the Instituto Nacional de Astronomia, \'Optica y
Electr\'onica (Puebla, Mexico) which provided a very stimulating and friendly atmosphere.  The organizers of this workshop are gratefully acknowledged. Thanks also  to Cathy Ramsbottom and Claire Hudson for giving me access to some atomic data. The comments of Enrique P\'erez and of the referee, Fabio Bresolin, are gratefully acknowledged.

\end{acknowledgements}

\clearpage

\appendix

\section{Atomic data}

In the  version of PHOTO used in this work, the atomic data to compute line emissivities have been updated with respect to the ones listed in Stasi\'nska \& Leitherer (1996).  The recombination line emissivities for HI  are from Storey \& Hummer (1995) with collisional excitation contribution using data from Anderson et al. (2002). The emissivities of  HeI lines are from Benjamin et al. (1999) and those of HeII lines are from Storey \& Hummer (1995).  For the remaining elements, the references are  given in  Table A.1. In this table, the reference Froese Fischer \& Tachiev (2004a) 	refers to the internet site http://hf8.vuse.vanderbilt.edu.  After  our   compilation of atomic data was performed, the paper Froese Fischer \& Tachiev (2004b) was published, giving transition probabilities of all ions of interest.

\begin{table} [!hbp]
\caption{Atomic data for line emission}
 
\begin{tabular}{lll}
\hline

ion		& 		collision strengths	 	& 		transition probabilities			 	\\
		&			 	&					 	\\
C I  		&		P\'equignot \& Aldrovandi (1976) 	 	&		Galavis et al. ( 1997a), Mendoza \& Zeippen (1999)		 	\\
C II & Wilson \&  Bell (2002) & Galavis et al. (1998) \\
C III		&		Berrington (1985, 1992), Burke et al. (1985),    	&		Tachiev \& Froese Fischer  (1999) 			 	\\
C IV		&		Griffin et al. (2000)Ê 	&		Kingston \& Hibbert (2001) \\
		&			 	&					 	\\
N I 		&		Berrington \& Burke (1981)   	 	&		Tachiev \& Froese Fischer (2002)	 	\\
N II		&		Hudson \&  Bell (2004)   	&		Galavis et al. (1997a), Storey \& Zeippen (2000)		 	\\
N III		&		Stafford et al.  (1994) 	 	&		Galavis et al. (1998) 	 	\\
N IV		&		Ramsbottom  et al. (1994) 	 	&		Tachiev \& Froese Fischer (1999)	 	\\
		&			 	&					 	\\
O I		&		Zatsarinny \& Tayal (2003)	 	&		Galavis et al. (1997) 	\\
O II		&		McLaughlin\& Bell (2000) 	 	&		Tachiev \& Froese Fischer (2002) 	\\
O III 		&		Aggarwal \& Keenan (1999)	 	&		Galavis et al.  (1997),  Mendoza et al. (1999),  Storey \& Zeippen (2000)
			 	\\
O IV		&		Zhang et al. (1994)	 	&		Galavis et al. (1997b) \\
O V     &    Zhang \& Sampson (1993) & Tachiev \& Froese Fischer (1999) \\
		&			 	&					 	\\
Ne II		&		Griffin et al.  (2001)  	 	&		    Froese Fischer \& Tachiev (2004a) 			 	\\
Ne III  		&		McLaughlin \& Bell (2000) 	&		Galavis et al.  (1997a),  Storey \& Zeippen (2000) 			 	\\
Ne IV		&		Ramsbottom et al. (1998) 	&		Merkelis et al. (1999)	 	\\
Ne V  		&		Griffin  \& Badnell (2000)	 	&		Galavis et al. (1997a) 	 	\\
Ne VI		&		Mitnik et al.  (2001) 	&		Galavis et al. (1998) 	\\
		&			 	&					 	\\
Mg II		&		Sigut \&  Pradhan (1995)	&		  Froese Fischer \& Tachiev (2004a) 		 	\\
Mg IV		&		Berrington et al. (1998)	 	&					 	\\
Mg V		&		Butler \& Zeippen (1994)	 	&		Tachiev \& Froese Fischer  (2002a)	 	\\
		&			 	&					 	\\
Si II		&		Pradhan (1995) 	&		 Froese Fischer \& Tachiev (2004a) 				 	\\
		&			 	&					 	\\
S II 		&		Ramsbottom  et al., (1996)	&		 Froese Fischer \& Tachiev (2004a) 				 	\\
S III		&		Tayal \& Gupta (1999)	 	&	Froese Fischer \& Tachiev (2004a) 				 	\\
S IV		&		Tayal (2000) 	&		Froese Fischer \& Tachiev (2004a) 				 	\\
		&			 	&					 	\\
Ar II		&		Pelan \& Berrington  (1995)	 	&		Mendoza (1983)	 	\\
Ar III		&		Galavis et al. (1998)	&		Mendoza \&  Zeippen (1982a) 	\\
Ar IV		&		Ramsbottom et al. (1997) + 2004 private	communication	&		Mendoza \&   Zeippen (1982b)	 	\\
Ar V		&		Galavis et al. (1995)	&		  Froese Fischer \& Tachiev (2004a) 				 	\\
Ar VI		&		Saraph \& Storey (1996) 	&	  Froese Fischer \& Tachiev (2004a) 			 	\\

\hline
\end{tabular}
\end{table}

\end{document}